\documentclass{acm_proc_article-sp}

\usepackage{amsfonts,amsmath}

\usepackage{latexsym}
\usepackage{epsfig}
\usepackage{theorem}

\newtheorem{theorem}{Theorem}
\newtheorem{definition}[theorem]{Definition}
\newtheorem{corollary}[theorem]{Corollary}
\newtheorem{lemma}[theorem]{Lemma}
\newtheorem{fact}[theorem]{Fact}
\newcommand{\sketchofproof}{\noindent {\em Sketch of proof}.~}
\newcommand{\eproof}{\hspace{\stretch{1}}$\Box$\\ }

\newlength{\pgmtab}          
\def\np{\mbox{\rm NP}}

\def\C{\mathcal{C}}

\def\P{\mathcal{ P}}

\def\maxlife{{\sc Max LifeTime}}

\def\DC{{\sc $l$-DISK COVER}}

\def\broadcast{{\sc Min Energy Broadcast}}
\def\lbroadcast{{\sc $l$-Min Energy Broadcast}}
\def\2dbroadcast{{\sc Min h-Euclidean Broadcast Subgraph}}
\def\3dbroadcast{{\sc Min 3D-Euclidean Broadcast Subgraph}}

\def\cellbr{\mbox{\sc {cell-alg}}}
\newcommand{\bro}{\mbox{\sc {Broadcast}}}
\def\conta{\mbox{\tt counter}}

\def\opt{\mbox{\tt opt}}

\def\dist{\mbox{\tt dist}}

\def\sF{{\cal F}}

\def\sC{{\cal C}}

\def\sR{{\cal R}}
\def\sZ{{\cal Z}}

\def\OCT{\mathcal{OCT}}

\def\Cnk#1#2{\left( \begin{array}{c} #1 \\ #2 \end{array} \right)}

\def\sR{{\mathcal R}}
\def\opt{{\sf opt}}
\def\mis{{\sf cost}}

\begin{document}


\title{ Minimum-energy broadcast
 in random-grid ad-hoc networks: approximation and distributed algorithms\titlenote{Research
 partially supported
    by the European Union under the Project \emph{AEOLUS} }
  }

\numberofauthors{5}

\author{
\alignauthor
Tiziana Calamoneri\\
\affaddr{Dipartimento di Informatica}\\
\affaddr{Sapienza Universit\`a di Roma}\\
\email{calamo@di.uniroma1.it}
\alignauthor
Andrea E.F. Clementi\\
\affaddr{Dipartimento di Matematica}\\
\affaddr{Universit\`a di Roma II, ``Tor Vergata''}\\
\email{clementi@mat.uniroma2.it}
\alignauthor
Angelo Monti\\
\affaddr{Dipartimento di Informatica}\\
\affaddr{Sapienza Universit\`a di Roma}\\
\email{monti@di.uniroma1.it}
\and
\alignauthor
Gianluca Rossi\\
\affaddr{Dipartimento di Matematica}\\
\affaddr{Universit\`a di Roma II, ``Tor Vergata''}\\
\email{gianluca.rossi@uniroma2.it}
\alignauthor
Riccardo Silvestri\\
\affaddr{Dipartimento di Informatica}\\
\affaddr{Sapienza Universit\`a di Roma}\\
\email{silver@di.uniroma1.it}
}

\maketitle


\begin{abstract}
The \broadcast\ problem consists in  assigning transmission ranges
to the nodes of an   ad-hoc   network in order to guarantee a
directed spanning tree from a given \emph{source} node and, at the
same time,  to minimize the  energy consumption (i.e. the
\emph{energy cost})  yielded by the range assignment.
 \broadcast\ is known to be \np-hard.
\\   We consider \emph{random-grid
networks} where nodes are chosen independently at random from
the  $n$ points of a $\sqrt n \times \sqrt n$ square grid in the
plane. The probability of the existence of a node at a given point
of the grid does depend on that point, that is, the probability
distribution can be \emph{non-uniform}. \\
    By using information-theoretic arguments,   we
          prove    a   lower bound   $(1-\epsilon)  \frac n{\pi}$
        on
     the energy cost of any feasible solution for this problem.
     Then, we provide
     an efficient solution of   energy cost not larger than
      $1.1204 \frac n{\pi}$. \\
   Finally, we   present   a fully-distributed
protocol that constructs a broadcast range assignment of   energy cost
not larger than $8n$,
  thus still yielding constant approximation.
  The \emph{energy load} is well balanced and, at the same time, the
    \emph{work complexity} (i.e. the energy due  to all message
    transmissions of the protocol)
     is  asymptotically  optimal.  The   completion time of
     the protocol is only an $O(\log n)$ factor slower  than the
     optimum. The   approximation quality of our distributed solution is also
experimentally evaluated. \\
 All bounds   hold with probability at least
$1-1/n^{\Theta(1)}$.
\end{abstract}

\section{Introduction}
 \paragraph{Range assignments  in ad-hoc networks}
  In ad-hoc networks,   nodes  are able to vary  their
   transmission ranges  in order
   to provide good network connectivity and low energy consumption at the same time.
More precisely, the   transmission ranges    determine a
(directed) \emph{communication graph} $G(S,E)$ over the set $S$ of
nodes: a node $v$, with transmission range $r$, can transmit to
another node $w$ (so, edge $(u,w) \in E$) if and only if $w$
belongs to the \emph{disk} centered in $v$ and of radius $r$. The
transmission range of a node depends, in turn, on the energy power
supplied to the node. In particular, the power $P_{v}$ required by
a node $v$ to correctly transmit data to another station $w$ must
satisfy the inequality
  $
  P_v \ge \dist(v,w)^{2}$,   where
 $\dist(v,w)$   is the Euclidean distance between $v$ and $w$.
  In several  works \cite{A05,CCPRV01,WieNguEph00,KirKraKriPel00},
   it is assumed that nodes can arbitrarily vary  their
  transmission range over the  ''large''  set $\{ \dist(s,t) | s,t \in S \}$.
   However, in some important   network scenarios   (like sensor networks),
   this assumption is not realistic: the adopted technology allows nodes  to have only \emph{few}
  possible transmission range values. For this reason, we
adopt the model considered in \cite{CD05,CWL06,EG01,WY04} where
nodes  are able to
 choose  their  transmission range   from a     \emph{restricted}
 set $\Gamma$.

\noindent A fundamental class of algorithmic problems arising from
  ad-hoc wireless networks consists in  the  \emph{range
assignment problems}: find a transmission range assignment $r:S
\rightarrow \Gamma$ such that (1) the resulting communication
graph satisfies a given connectivity property $\Pi$, and (2) the
\emph{energy cost} $\mis(r) = \sum r(s)^2$   of the assignment  is
minimized (see~\cite{WieNguEph00,KirKraKriPel00}). \\
 Several research works
\cite{A05,CCPRV01,CFM07,WieNguEph00} have been devoted to the \broadcast\  problem
where  $\Pi$ is defined as follows: \emph{Given a  source node
$s$, the communication graph has to contain a directed spanning
tree rooted at $s$ (a branching from $s$)}.
 Previous theoretical  results on   \broadcast\ concern
\emph{worst-case} analysis only. This problem is known to be
\np-hard \cite{CCPRV01} even when $\Gamma =  \{0, l_1 , l_2 \}$
for  $l_1 < l_2$ and $l_1$ is set to any fixed  positive constant.
The most famous approximation algorithm is the \emph{MST}-based
heuristic \cite{WieNguEph00}. This heuristics works in
$\Theta(n^2)$ time and  its performance analysis has been the
subject of several works over the last years
\cite{CCPRV01,FKNP07,WCLF01}. In \cite{A05}, it is finally proved
the \emph{tight}   bound 6 on its approximation ratio. More
recently, a new polynomial-time algorithm is provided in
\cite{CFM07} that achieves approximation ratio close to 4. This
algorithm applies  a rather complex edge-contraction technique on
the MST-based solution.   Its present best version works in
$\Theta(n^5)$ time and the design of any efficient distributed
version seems to be a very hard task.

\noindent It is important to observe that the MST-based heuristic
is ''far''
  from achieving  optimal solutions even   on a complete square
\emph{grid}  of $n$ points  \cite{CCDLMS06,FNP05}: its worst-case
approximation  ratio on such  grids is not smaller than 3. In
\cite{FNP05}, it is also experimentally observed that this
heuristic has  a  bad behavior when applied to random
\emph{regular} instances such as \emph{faulty} square grids.
Furthermore, the MST-based heuristic requires  a large range set
$\Gamma$.

\noindent  The above discussion leads us to study   \broadcast\
over \emph{random grid} networks.
    Given a  $\sqrt n \times \sqrt n$ grid of points of the Euclidean
plane (without loss of generality, adjacent points are placed at
unit distance),  each point\footnote{For the sake of simplicity,
we here assume
 that points are labelled with   index $i=1, \ldots , n$.}
$i$ is selected as a \emph{node} of the random grid network
independently
 with  probability $p_i$. This \emph{node probability}     can be any value in  the interval
 $[p_{min},p_{max}]$ where
  $p_{min}$ and $p_{max}$ are two arbitrary
 positive constants in the interval $(0,1)$.    We remark that our
  random grids    are in general   non uniform:
 Random grids   provide a good model for  several ad-hoc and
 sensor networks. On one hand,  by varying  the $p_i$'s values,     it is   possible
 to model non homogenous input configurations with regions of different node
 densities.  On the other hand, the grid structure guarantees  a minimal
 distance among nodes: this is often a desired   property in order to
 optimize area coverage and  avoid  message collisions.
 Nevertheless, as discussed later, all our results  also hold   for the
 standard  \emph{uniform random distribution} (i.e. the random input formed by choosing $n$ points
independently
  and uniformly at random from a  2-dimensional square)
\cite{KLSS04,Penrose}.

\paragraph{Our results}
 \textbf{$1)$} We
provide a lower bound on the energy cost of  feasible solutions
for \emph{any} range assignment problem on random grids where the
required property $\Pi$ \emph{implies} the existence of a
disk cover. We say that a range assignment is a
\emph{(disk) covering} assignment    if it guarantees that every node of
the network is within the \emph{positive} range of some node.
\broadcast\ is just one of those important cases requiring
covering range assignments. \\  Let      $l_1$ be the minimum
positive range in
 $\Gamma$. For any $0<\epsilon <1$, if  $l_1 = \Omega(\frac{1}{\epsilon})$
 then
  we prove that
      the   energy cost of any covering range assignment  is with high
       probability\footnote{Here and in the sequel the term {\em with high probability}
        means that the event holds with probability
at least $1 - \frac{1}{n^{a}}$ for some constant $a>0$.} (in
short, w.h.p.)
      at least $(1-\epsilon)\frac{n}{\pi}$.   Observe that the
      lower bound tends to $n/\pi$   for   any $l_1$   such that
      $l_1=\omega(1)$, so for minimal ranges much smaller
      than
           the \emph{connectivity threshold} $\Theta(\sqrt{\log n})$
           \cite{DP06,GK99,Penrose,SB03}.
\\  The  proof's technique  of the lower bound  departs
significantly from all those adopted in this  topic and uses
information-theoretic arguments. By using this  result, we will
prove that   the next two   algorithms are almost optimal.

 \noindent
       \textbf{$2)$}
We   provide a simple and efficient
   algorithm
for random grids that uses  minimal range $l_1 = \Theta(\sqrt{\log
n})$ and  returns a solution of energy cost not larger than $1.1204
\frac n{\pi}$ w.h.p.: In virtue of  our lower bound, this is very close to
the optimum.  Observe that  our lower bound  holds for any
covering range assignment while the upper bound holds for feasible
range assignments of \broadcast : this implies that, for $l_1 =
\omega(1)$, the extra-cost,   due to the required tree
connectivity property, is ''almost'' negligible in random grids
(it is still an open problem whether this is in fact negligible).
 Our algorithmic  solution works in $O(n\log
n)$ time and needs a set $\Gamma$ of \emph{logarithmic} size (in
$n$). The range assignment is inspired  to the one  provided in
\cite{CCDLMS06} for complete square grids (i.e.  every  point of
the grid is a node of the network). However,   the
probabilistic cost analysis of our construction   for \emph{random} grids is  definitely not related
to that   in \cite{CCDLMS06}.

   \noindent
       \textbf{$3)$} It is common opinion
       that the development of
       efficient, provably-good  \emph{distributed} algorithms is presently the major challenge about
        range assignment problems \cite{CHE02,WCLF01,WieNguEph00}.
          We provide an efficient distributed algorithm
        for  \broadcast\  on random grids. We investigate   the performance of the protocol
         in two different scenarios:
         \emph{single broadcast }  and  \emph{many-broadcast}, i.e., a sequence
         of consecutive broadcast operations.
        In both cases, besides the energy cost of the returned  range assignment,
        we consider further  important complexity aspects  that determine  the quality of a
        distributed solution.
\\ \emph{-  Work Complexity.}
          In  the ad-hoc network model, the \emph{work complexity}  of a distributed
       algorithm (i.e. protocol)
       for \broadcast\  is defined as the sum of   the energy cost
        of \emph{all}
       transmissions made by the protocol to perform the broadcast
       operation \cite{KPK08,LCW02,RR00}.
       This  complexity measure thus  considers both the   cost
       \emph{to construct} the range assignment and the cost \emph{to
       use} it to broadcast the message (the latter being
       exactly the cost of the range assignment  defined for    centralized algorithms).
        Since \emph{both} the above energy costs  are paid by the
       nodes,
       a protocol     can be really considered   energy efficient only if it has
          a small work complexity. \\
  \emph{- Energy-Load Balancing.} In some real ad-hoc
            networks (such as sensor networks), it is also important to equally distribute
            the energy load to all nodes. For instance,
            solutions, assigning large ranges to few nodes, are
              not feasible in scenarios where nodes have limited
              battery charges. In such a case, we aim to design
              solutions that are well energy-load balanced.
                Notice that, in the many-broadcast scenario, this
              corresponds to maximize \emph{network lifetime} according
               to the model in \cite{CCFS07,CKOZ06}. \\
\emph{- (Amortized) Completion Time.} Another relevant  aspect of
            a broadcast protocol is the  \emph{completion time}, i.e.,  the
            number of time steps required to complete one
            broadcast operation. In the many-broadcast scenario,       the
            \emph{amortized completion time} is the \emph{average} completion
            time for one   broadcast operation.

\noindent Our aim is to derive a protocol having
\emph{provably-good} performance with respect to \emph{all}  the
above   complexity aspects. To the best of our knowledge, no
 available protocol   has been shown to have
this overall performance. \\  We  first define  a very simple
range assignment where only one positive range  in $\Gamma$  is
used,  provided that it is not smaller than $c \sqrt{\log n}$
 where $c$ is a suitable positive constant (observe again that this value is asymptotically equivalent
 to the connectivity threshold).
  This
solution is then shown to be w.h.p. feasible and to have an energy
cost not larger than $8n$.  Thanks to our lower bound, the   achieved energy cost yields a constant
approximation ratio.  Moreover,    this simple range assignment  can
be constructed and managed by an efficient
 protocol. We assume  every node
initially knows $n$ and its relative position with respect to  the grid
only. Positioning information can be obtained
 by using GPS systems or Ad-Hoc Positioning System (APS) \cite{NN01}.
 This  assumption is reasonable in
\emph{static}  ad-hoc networks since   every node can
  store  \emph{once and for all} its position in the set-up phase.
  The protocol exploits  a
fully-distributed pivot-election strategy borrowed from
\cite{CCFS07}. \\
  We prove
  that the  work complexity  of the protocol is
equivalent to the energy cost of the centralized version and
hence, thanks again   to our lower bound
(clearly, a lower bound for the energy cost is also a lower bound for the work complexity),  it achieves  a constant
approximation ratio as well. It is important to emphasize that the
best distributed
       algorithm to compute an MST
       in the ad-hoc model has an
       expected work complexity  $\Omega(n\log n)$  \cite{GHS83,KPK08,L03,LWSF04}. By
       comparing this bound with the $\Theta(n)$ work complexity  achieved by our protocol, we can
       state that any MST-based solution  \cite{WieNguEph00,CFM07}     cannot yield
       good work complexity in this scenario.  Other distributed solutions have been
       considered in the literature \cite{CHE02,LCW02,RR00,WLF04}, however their
       performance analysis is based on experimental  tests only.
We also compared the work complexity of our protocol to the energy cost
of the centralized MST-based solution over thousands of random instances with
different sizes and densities. The average performance ratio
between the two solutions is always between 2 and 3 (see Section
\ref{sec::experiments}) thus confirming our analytical results. \\ Our protocol yields a
good energy-load balanced solution:  there are $\Theta(n/\log n)$
\emph{pivots}, i.e., the nodes    having  range $\Theta(\sqrt{\log
n})$ (the remaining nodes have range 0). Furthermore, thanks to
the pivot-election strategy \cite{CCFS07}, a good energy-load
balance is also obtained  with respect to an arbitrary sequence of broadcast
operations, i.e., for the many-broadcast scenario. At every new
operation, the pivot task is indeed assigned to nodes according to
a Round Robin rule. We show this yields an almost optimal
\emph{life-time} of the network according to the energy
consumption model in \cite{CCFS07,CKOZ06}. \\  As for the single
broadcast scenario,
            the completion time of our protocol is slower than the optimum
              by a $\Theta(\log n)$ factor. As for the many-broadcast scenario,
               when the number of broadcast operations
is $\Omega(\log n)$,  then the amortized completion time is
optimal.

\noindent Finally, we notice that,  by using  the   technique in \cite{LN06},  our
protocol can be emulated      on  the  standard  uniform random
distribution \cite{KLSS04,Penrose}.
   The same holds for the centralized algorithm achieving cost $1.1204
\frac n{\pi}$  and for the lower bound as well.
  The relative  proofs for the uniform distribution
  are easier and, so, they   are  omitted in this extended abstract.

\paragraph{Paper's Organization}  In Section \ref{sec::low}, we provide
the proof of
   the lower bound. In Section \ref{apx::uppbound}, we describe the
 centralized algorithm yielding almost optimal cost. Finally,
 Section \ref{sec::distrib} is devoted to the description of our
 distributed protocol and its analysis.

\subsection{Preliminaries}

 The square grid $R$ of $\sqrt n
\times \sqrt n$ points   will be indexed from $1$ to $n$. Without
loss of generality, the distance between adjacent points is set to
1. To each point $i$ of $R$, a   probability value $p_i$ is
assigned such that $p_{min} \le  p_i \le p_{max}$ where $p_{min}$
and $p_{max}$ are arbitrary constants in $(0,1)$. We consider the
random input model $\sR (R,p_1,\ldots ,p_n)$ where an instance
$S\subseteq R$ has probability

\[ P_R(S) = \prod_{i\in S} p_i \cdot   \prod_{i \in R - S} (1-p_i) \]

\noindent Observe that this probability distribution is equivalent
to select each point $i \in R$ independently with probability
$p_i$. A selected point will be called \emph{node}. In the sequel,
a subset $S\subseteq R$ selected according to  the above random
distribution will be simply called  \emph{random set}.

\noindent
A set of disks $ \C =  \{ D_1, \ldots ,  D_m \}$ is said
 to be  an $l$-cover for $S\subseteq R$ if the following
properties hold: \emph{i)} All disks of $\sC$ have radius at
least $l$, where $l$
    is some positive value.
   \emph{ii)} Every disk of $\sC$ has its center on a node of $S$.
   \emph{iii)}  Every node of $S$ is covered by some disk of $\sC$.

\noindent Observe that  a range assignment $r:S \rightarrow
\Gamma= \{0, l_1 , l_2\ldots,   l_k \}$ can be represented by the
family of disks ${\cal C} = \{C_1, \ldots , C_{\ell} \}$ yielded
by    the positive values of $r$, and its   energy cost   is
defined as

\begin{equation}\label{eq::racost}
  \mis({\cal C}) =  \sum_{i = 1}^{\ell} {r_i^2}  \; \mbox{ where }
{r_i} \mbox { is the radius of } {C_i}
\end{equation}

\noindent
 Furthermore, a feasible range assignment for the
\broadcast\ problem, with input $S$ and $\Gamma$, uniquely
determines an $l_1$-cover for $S$ having the same cost. Notice
that the converse is not true in general.

\section{The lower bound} \label{sec::low}

In this section, we provide a lower bound on the cost of any
\emph{covering} range assignment for a random set  $S\subseteq R$.

\begin{definition} Let  ${\bf Pr}[R,\epsilon, l]$
 be the probability that a random set $S\subseteq R$  has
 an  $l$-cover  of cost not larger than  $(1-\epsilon) \frac n {\pi}$.
 \end{definition}
\begin{theorem}\label{t2}
Let $\delta, p_{min}$ and $ p_{max}$ be three   constants such
that  $0<\delta<1$ and  $0< p_{min}\leq p_{max}<1$. Let
$S\subseteq R$ be any random set.  Then, for any $\epsilon$ with
$0<\epsilon<1$, for   sufficiently large $n$, and for \[  l \ge
\frac{5(1-\epsilon)^{\frac{1}{6}}}{\epsilon(1-\delta) p_{min}}
\] it holds that
\[ {\bf Pr}[R,\epsilon, l] \leq  2^{-\frac{1}{100}(1-\delta)\epsilon p_{min} n + \log(4n) + t}
+ e^{
-\frac{\delta^2}{2}p_{min}\left\lfloor\frac{n}{t}\right\rfloor+ t}
\]
\[ \mbox{where  } \ t=\left\lceil \frac{100}{3}\frac{\log
\left(\frac{1-p_{min}}{p_{min}}\frac{p_{max}}{1-p_{max}}\right)+1}{
\epsilon(1-\delta) p_{min}} \right\rceil \]
\end{theorem}

\noindent The above theorem  clearly implies our lower bound
stated in the Introduction and it requires no restriction about
the transmission-range set $\Gamma$ but a lower bound on $l_1=l$
that does not depend on $n$. In particular, if $\epsilon$ is
\emph{any} positive constant then,
 for sufficiently large grids and a sufficiently large \emph{constant} $l$ (so $l$ does not depend on $n$),
${\bf Pr}[R,\epsilon, l]$ is not larger than the inverse of an
exponential function in $n$.

\noindent The theorem's proof
     makes use of the following  combinatorial result.

\begin{lemma}\label{1}
Let $R_1, \ldots, R_t$ be a partition of the $n$ points in $R$ and
let $(k_1, \ldots, k_t)$ be any $t$-tuple of  integers such that
$0\leq k_j\leq |R_j|=n_j$. Then, the number of subsets $S$ of $R$
such that $|S\cap R_j|=k_j$ ($1 \le j \le t$)      admitting
an $l$-cover $\sC$ with $l\geq \sqrt{e}$ and
  $ \mis(\sC)\leq \frac{(1-\epsilon)n}{\pi}$ is at most
\[2^{\lambda(n,q, \epsilon,l,t) } \cdot \prod_{j =1
}^{t}\Cnk{n_j}{k_j} \] where $q=\min\left\{\frac{k_j}{n_j}|1\leq
j\leq t\right\}$  and \quad $\lambda(n,q, \epsilon,l,t) =$
\begin{small}
\[ \left(-q\log e
\left(1-(1-\epsilon)\left(1+\frac{1}{2l^2}+\frac{1}{\sqrt{2}l}\right)^2\right)+
\frac{1-\epsilon}{\pi l^2}\log \frac{64 e \pi
l^6}{1-\epsilon}\right) n \] \[ + \log (4n)+t\] \end{small}
\end{lemma}

\noindent    We  now  provide a brief description of  the
\emph{information-theoretic} approach adopted to prove  the above
lemma.

  \noindent
  Let $S$ be a subset of points of $R$ satisfying the
hypothesis of the lemma. By exploiting the $l$-cover  $\sC$, we
will  prove that $S$ can be encoded into  a binary string $cod(S)$
of length at most
 \[\log\left(
2^{\lambda(n,q, \epsilon,l,t) } \cdot \prod_{j =1
}^{t}\Cnk{n_j}{k_j}\right) \] The lemma thus follows since  the
number of these sets $S$   cannot exceed the number of binary
strings of the above length. \\

\paragraph{Proof of Lemma \ref{1}}
Let $S$ be a subset of points of $R$ satisfying the hypothesis of
the lemma.   Consider the $l$-cover $\sC'$ of $S$ having the same
centers of $\sC$ and where each radius $r$ in $\sC$ is replaced
with a radius $r'=\sqrt{\lceil r^2 \rceil}$. Clearly, this change
is negligible in terms of cost.

\noindent We now show that, thanks to $\sC'$, $S$ can be encoded
into  a binary string of length at most
 \[\log\left(
2^{\lambda(n,q, \epsilon,l,t) } \cdot \prod_{j =1
}^{t}\Cnk{n_j}{k_j}\right) \] Thus the thesis follows by noting
that the number of these sets $S$   cannot exceed the number of
binary strings of the above length.

\noindent The binary string $cod(S)$ encoding  $S$ is the
concatenation of four substrings $NUM$, $CEN$, $RAD$ and $COV$.

\begin{description}
\item[a)] $NUM$ reports the number $m$ of centers of    $\sC$.

\item[b)] $CEN$ reports information to recover  the indices of the
$m$ nodes of $S$ that are centers in $\sC$ (we assume that the $n$
points in the grid are numbered from $1$ to $n$).

 \item[c)] $RAD$ reports information to recover the radii of the $m$
nodes in $\sC'$.

 \item[d)] $COV$ reports information to recover the indices of the
nodes in $S$.
\end{description}

\noindent We now explain how these data are encoded and then bound
the length of each of the four substrings.

\begin{description}
  \item[a)] The number $m$ of centers in $\sC$ is at most $|S|\leq
n$. Thus we encode it by a binary string of fixed length (i.e.
$\lceil\log (n+1)\rceil$). Hence

\begin{equation}\label{NUM}
|NUM|=\lceil\log (n+1)\rceil \leq \log n+1
\end{equation}

\item[b)]  The centers of $\sC$ are a subset of the $n$ points in
$R$ and so we
 encode  them by a string of fixed length, i.e.

 \[\left\lceil\log \left(\left(\begin{array}{c}
n \\m
\end{array}\right)+1\right)\right\rceil \]

\noindent Since in the cover, each of the $m$ centers has radius
at least $l$,
 it must hold $ml^2\leq \mis(C)$.    From the hypothesis   $\mis(C) \le (1-\epsilon)\pi n$,
 we
 get
\begin{equation}\label{m}
m\leq \frac{(1-\epsilon)n}{\pi l^2}
\end{equation}
As for   substring $CEN$,   we obtain
\begin{eqnarray}
|CEN| &= & \left\lceil \log \binom{n}{m} +1\right\rceil
\leq \log \binom{n}{m}+1\nonumber\\
&\leq& m \log \frac{\epsilon n}{m}+1\nonumber
 \leq \frac{(1-\epsilon)n}{\pi l^2}\log \frac{e\pi
l^2}{1-\epsilon} +1\label{CEN}
\end{eqnarray}
Observe  that in the above inequalities   we used
\[ \left(\frac{en}{m}\right)^m \le \left(\frac{e\pi
l^2}{1-\epsilon}\right)^{\frac{(1-\epsilon)n}{\pi l^2}} \] since
  the function $\left(\frac{\epsilon n}{x}\right)^x$ is increasing
in the range $[1,n]$; then, from (\ref{m}),  $m$ is in the range
$[1,\frac{(1-\epsilon)n}{\pi l^2}]$.

  \item[c)]
  Let now $r'_{1}, r'_{2},\ldots r'_{m}$ be the
 radii in $\sC'$ arranged by increasing order of the indices of the $m$ centers. In order to
  give the
 information on the radii of   $\sC'$, we encode     string
  $\lceil r_{1}\rceil^2\# \lceil r_{2}\rceil ^2\#\ldots \lceil r_{m}\rceil^2\#$  in binary where
   bit $0$ is encoded as $00$,
      bit $1$ as $11$ and the symbol $\#$ as $01$. We thus get
\begin{small}
 \begin{eqnarray}
|RAD| & = & 2m + 2\sum_{i=1}^{m}\lceil\log (\lceil{r_i}^2\rceil
+1)\rceil\nonumber\\
&\leq& 2m + 2\sum_{i=1}^{m}\lceil\log ({r_i}^2
+2)\rceil\nonumber\\
&\leq&
2m + 2\sum_{i=1}^{m}(\log {r_i}^2+2)\nonumber \\
& =& 2\log \prod_{i=1}^{m}{r_i}^2 +6m
 \leq   2\log {\left(\frac{cost(C)}{m}\right)}^m  +6m \nonumber\\
 &\leq&  2\log{l}^{2\frac{\mis(C)}{l^2}}  +6m \nonumber
 \leq 4\frac{(1-\epsilon)n} {\pi l^2}\log {l}
+6\frac{(1-\epsilon)n}{\pi l^2}\\
 &=&  4\frac{(1-\epsilon)n}{\pi l^2}\log {(\sqrt{8}l)}\label{RAD}
\end{eqnarray} \end{small}
In the above inequalities, we    first used
$\prod_{i=1}^{m}{r_i}^2 \le {\left(\frac{\mis(C)}{m}\right)}^m$
since  the product is maximized when all   factors have the same
value. Next, we used  ${\left(\frac{\mis(C)}{m}\right)}^m \le
{l}^{2\frac{\mis(C)}{l^2}}$ since the function
${\left(\frac{\mis(C)}{x}\right)}^x$ is increasing in the range
$[1,\frac{\mis(C)}{e}]$; the value  of $m$ is in the range
$[1,\mis(C)/l^2]$; $ml^2\leq \mis(C)$ and $\mis(C)/l^2\leq
 \mis(C)/e$ for $l\geq \sqrt{e}$. Finally, we bounded $m$
using (\ref{m}).

\item[d)] In order to encode  the nodes in $S$, we use $t$
strings. The $j$-th string encodes the $k_j$ points of $S$ in
$R_j$. The $k_j$ points of $S$ in $R_j$ covered by $\sC'$  are a
subset of the $n'_j$ points in $R_j$ covered by $\sC'$. Hence, we
encode these $k_j$ points  with a binary string of length

\[\left\lceil\log \displaystyle\prod_{j=1}^{t}\Cnk{n'_j}{k_j}
+1\right\rceil \]

\noindent  Since for integers $a,b$ and $c$ where $a\leq b\leq c$
it holds that $\Cnk{b}{a}\leq \left(\frac{b}{c}\right)^a
\Cnk{c}{a}$, we get

\begin{small}
\begin{eqnarray} |COV| & = &  \sum_{j=1}^{t}\left\lceil\log
\left( \binom{n'_i}{k_i} +1\right) \right\rceil  \leq
 \sum_{j=1}^{t}\left(\log  \binom{n'_i}{k_i} +1 \right)  \nonumber \\
 & \leq &  \log \prod_{j=1}^{t}\binom{n'_i}{k_i} +t
 \leq    \log  \left(\prod_{j=1}^{t}\left(\frac{n'_j}{n_j}\right)^{k_j}\binom{n_j}{k_j}\right)
 +t \nonumber\\
   & \le &  \log  \prod_{j=1}^{t}\left(\frac{n'_j}{n_j}\right)^{k_j}+ \log\prod_{j=1}^{t}\binom{n_j}{k_j} +t \nonumber\\
 & = & \log \prod_{j=1}^{t}\left(1-\frac{n_j-n'_j}{n_j}\right)^{k_j}+
     \log\prod_{j=1}^{t}\binom{n_j}{k_j} +t \nonumber\\
  & \leq &  \log  \prod_{j=1}^{t}e^{-\frac{k_j}{n_j}(n_j-n'_j)}+ \log\prod_{j=1}^{t}\binom{n_j}{k_j} +t\nonumber\\
  & \leq & \log  e^{-\sum_{j=1}^{t}q(n_j-n'_j)}+ \log
  \prod_{j=1}^{t}\binom{n_j}{k_j} +t
\nonumber\\
 & = & -q(n-n')\log e + \log\prod_{j=1}^{t}\binom{n_j}{k_j} +t
\label{COV1}
\end{eqnarray}
\end{small}

\noindent where $n'$ is the number of points of $R$ covered by
$\sC'$.  We now give an upper  bound for  $n'$.

\noindent Let $x$ be the number of points of $R$ covered by a disk
$D$ of radius $r$ and, for each of these points, consider the
square of area $1$ centered in the point. These $t$ squares are
disjoint and are covered by the disk obtained by extending the
radius $r$ of   disk $D$ to $r+ 1/\sqrt{2}$ .  So,
 the number of   points of   $R$ covered by a disk of radius $r$ is bounded by
$\pi (r + 1/\sqrt{2} )^2$. Moreover, it holds that
\[r'_i=\sqrt{\lceil{r_i}^2\rceil}\leq \sqrt{{r_i}^2+1}< r_i
+\frac{1}{2l}\]
 where the last inequality follows since $r_i\geq l$. We thus
 obtain
\begin{small} \begin{eqnarray*}
n'&\leq& \sum_{i=1}^{m}\pi\left(r'_i +\frac{1}{\sqrt{2}}\right)^2
\leq \sum_{i=1}^{m}\pi\left(r_i
+\frac{1}{2l}+\frac{1}{\sqrt{2}}\right)^2\\
&=& \pi\left( \sum_{i=1}^{m}{r_i} ^2+
\left(\frac{1}{2l}+\frac{1}{\sqrt{2}} \right)^2m +
2\left(\frac{1}{2l}+\frac{1}{\sqrt{2}} \right)\sum_{i=1}^{m}r_i
\right)
\end{eqnarray*}
\end{small}

 We can now use  H$\ddot{o}$lder's inequality  and obtain
\[
\sum_{i=1}^{m}{r_i}=\sum_{i=1}^{m}\left({r_i}^2\right)^\frac{1}{2}
\leq
m\left(\frac{\sum_{i=1}^{m}{r_i}^2}{m}\right)^\frac{1}{2}=\sqrt{m\cdot
\mis(C)}\]   Since $m\leq  \mis(C)/l^2$, we get
\begin{small} \begin{eqnarray}\label{totale}
n' & \leq & \pi \mis(C)\left( 1 +
\left(\frac{1}{2l}+\frac{1}{\sqrt{2}} \right)^2\frac{1}{l^2} +2
\left(\frac{1}{2l}+\frac{1}{\sqrt{2}}
\right)\frac{1}{l}\right)\nonumber\\
&\leq&
(1-\epsilon)\left(1+\frac{1}{2l^2}+\frac{1}{\sqrt{2}l}\right)^2n
 \nonumber
\end{eqnarray} \end{small}
From (\ref{COV1}), we get
\begin{small} \begin{eqnarray}
|COV| & \leq & -q\log
e\left(1-(1-\epsilon)\left(1+\frac{1}{2l^2}+\frac{1}{\sqrt{2}l}\right)^2\right)n
\nonumber\\
&&+ \log\prod_{j=1}^{t}\Cnk{n'_j}{k_j} +t\label{COV}
\end{eqnarray} \end{small}
\end{description}

 \noindent
  We now combine   the bounds in (\ref{NUM}), (\ref{CEN}), (\ref{RAD}) and (\ref{COV})
  (respectively on the lengths of NUM, CEN, RAD and COV)
   and      obtain
\begin{small} \begin{eqnarray*}
|cod(S)| & \leq &
 \log n + 1+ \frac{(1-\epsilon)n}{\pi l^2}\log \frac{e \pi
   l^2}{1-\epsilon} +1\\
& + & 4\frac{(1-\epsilon)n}{\pi l^2}\log (\sqrt{8}l) + \log\prod_{j=1}^{t}\Cnk{n'_j}{k_j} +t\\
& -&  q\log
e\left(1-(1-\epsilon)\left(1+\frac{1}{2l^2}+\frac{1}{\sqrt{2}l}\right)^2\right)n\\
& = &   \log(4n) + \log \prod_{i =1 }^{t}\binom{n_i}{k_i}+t +
\left( \frac{1-\epsilon}{\pi l^2}\log \frac{64 e \pi
l^6}{1-\epsilon} \right.\\
& &\left. - q\log
e\left(1-(1-\epsilon)\left(1+\frac{1}{2l^2}+\frac{1}{\sqrt{2}l}\right)^2\right)\right)n
\end{eqnarray*} \end{small}
\eproof
\paragraph{Proof of Theorem \ref{t2}}
We  assume  \[ n\geq \frac{1000}{9}\left(\frac{\log
\left(\frac{1-p_{min}}{p_{min}}\frac{p_{max}}{1-p_{max}}\right)+1}{\delta
(1-\delta) p_{min}\epsilon}\right)^2  \]
 For any $S$, with $S\subseteq R$, define the binary function $\chi$ as follows
\[\chi_{\epsilon, l}(S)=\left\{ \begin{array}{ll}1 & \mbox{if $S$ has an $l$-cover
 of cost at most $ (1-\epsilon)n/ \pi$} \\
0 & \mbox{otherwise}\end{array}\right. \]

\noindent Clearly, it holds that
\begin{eqnarray}\label{uno}
{\bf Pr}[R,\epsilon, l] = \sum_{S\subseteq R}P_R(S)\chi_{\epsilon,
l}(S)
\end{eqnarray}
 Let us partition  $R$ into  $t$ regions $R_1,R_2,
\ldots, R_t$ where  $|R_j|=n_j$ such that for $1\leq j\leq t$
 \[ \left\lfloor
\frac{n}{t}\right\rfloor\leq n_j\leq \left\lceil
\frac{n}{t}\right\rceil \mbox{ and }
 \ R_j=\left\{\sum_{i=1}^{j-1}n_i +k | 1\leq k \leq n_j\right\}.
 \]
Define $\mu_j$ as the expected number of points in $j$, i.e.,
$\mu_j=\sum_{i\in R_j}p_i$. Let $\sF$ be the family of subsets of
$R$ having, in each region, a number of points not too small
w.r.t. the expected number, i.e, \[\sF=\{S\in 2^R \;|\; |S\cap
R_j| \geq (1-\delta)\mu_j, 1\leq j\leq t\}\]
 From (\ref{uno})  we get
 \begin{eqnarray}\label{due}
{\bf Pr}[R,\epsilon, l] =
 \sum_{S\subseteq  \sF}P_R(S)\chi_{\epsilon, l}(S)+
 \sum_{S\subseteq 2^R -\sF}P_R(S)\chi_{\epsilon, l}(S)
\end{eqnarray}
 We start   giving an upper bound  on  the first addend
of  the right-hand of the above equation.
 Let
 \begin{multline*}
   A=\{\vec{k}=(k_1,\ldots , k_t)\;|\; \vec{k}\in \sZ^t \\
   \mbox{and}\ (1-\delta)\mu_j\leq k_j \leq n_j, 1\leq j\leq t\}
 \end{multline*}
and, for each $\vec{k} \in A$, define
\[\sF_{\vec{k}}=\{S\in \sF \;|\; |S\cap
R_j| = k_j, 1\leq j\leq t\} \] Consider any set  $S_{\vec{k}} \in
 \sF_{\vec{k}}$ such that $P_R(S_{\vec{k}})\geq P_R(S)$ for every
$S \in     \sF_{\vec{k}}$. Then,
\begin{small} \begin{multline} \sum_{S\subseteq
\sF}P_R(S)\chi_{\epsilon, l}(S)  =    \sum_{\vec{k} \in
A}\sum_{S\in \sF_{\vec{k}}} P_R(S)\chi_{\epsilon, l}(S)
 \leq\\   \sum_{\vec{k} \in A}\sum_{S\in \sF_{\vec{k}}}P_R(S_{\vec{k}})
 \chi_{\epsilon, l}(S)
=  \sum_{\vec{k} \in A}P_R(S_{\vec{k}})\sum_{S\in
\sF_{\vec{k}}}\chi_{\epsilon, l}(S)\\  \leq  \sum_{\vec{k} \in
A}P_R(S_{\vec{k}})
2^{\lambda\left(n,\min\left\{\frac{k_j}{n_j}|1\leq j\leq t\right\}
, \epsilon, l,t\right)
}\displaystyle\prod_{i=1}^{t}\left(\begin{array}{c} n_i \\ k_i
\end{array}\right)\label{primo1}
\end{multline} \end{small}
  where the last step follows from Lemma \ref{1}. Function
$\lambda$ is decreasing in $q$ and
\begin{small}\begin{multline*}
\min_{1\leq j\leq t}\left\{\frac{k_j}{n_j}\right\} \geq
\min_{1\leq j\leq t}\left\{\frac{(1 -\delta)\mu_j}{n_j}\right\}
\geq
\min_{1\leq j\leq t}\left\{\frac{(1-\delta)p_{min}n_j}{n_j}\right\}\\
=
 (1-\delta)p_{min}
\end{multline*}
\end{small}
  We thus      get
 \begin{small} \begin{multline} \sum_{S\subseteq
\sF}P_R(S)\chi_{\epsilon, l}(S)  \leq  \sum_{\vec{k} \in
A}P_R(S_{\vec{k}}) 2^{\lambda(n,(1-\delta)p_{min}, \epsilon, l,t)
}\displaystyle\prod_{i=1}^{t}\binom{n_i}{k_i} =\\=
  2^{\lambda(n,(1-\delta)p_{min}, \epsilon, l,t)}\sum_{\vec{k}
\in A}P_R(S_{\vec{k}})
\displaystyle\prod_{i=1}^{t}\binom{n_i}{k_i} =\\ =
2^{\lambda(n,(1-\delta)p_{min}, \epsilon, l,t)}\sum_{\vec{k} \in
A}\sum_{S\in \sF_{\vec{k}}}P_R(S_{\vec{k}}) \label{secondo2}
\end{multline} \end{small}
Assume without loss of generality   that the points in $R$ are
numbered in increasing order w.r.t. their probability i.e.
$0<p_{min}=p_1\leq p_2\leq \ldots p_n=p_{max}<1$. Let
$q_0=p_{min}\leq q_1\leq q_2 \ldots \leq q_t=p_{max}$ where
$q_j=\max\{p_i | i\in R_j\}$, for $1\leq j\leq t$.  Consider any
set  $S\in  \sF_{\vec{k}}$, then

\begin{small}
\begin{multline} P_R(S_{\vec{k}}) =
\prod_{j=1}^{t}P_{R_j}(S_{\vec{k}}\cap R_j)\\
 =  \prod_{j=1}^{t}\left(\prod_{i\in S_{\vec{k}}\cap R_j }p_i  \prod_{i\in R_j -S_{\vec{k}}}
  (1-p_i)\right)
 = \prod_{j=1}^{t}\left( \prod_{i\in S\cap R_j }p_i  \right.\\ \left. \prod_{i\in
R_j -S}(1-p_i) \prod_{i\in (S-S_{\vec{k}})\cap R_j
}\frac{1-p_i}{p_i}\prod_{i\in (S_{\vec{k}}-S)
\cap R_j}\frac{p_i}{1-p_i}\right)\nonumber \\
\leq \prod_{j=1}^{t}\left( P_{R_j}(S\cap R_j) \prod_{i\in
(S-S_{\vec{k}})\cap R_j }\frac{1-q_{j-1}}{q_{j-1}}\prod_{i\in
(S_{\vec{k}}-S)\cap R_j}\frac{q_j}{1-q_j}\right)
\\
 = \prod_{j=1}^{t}\left( P_{R_j}(S\cap R_j)
\left(\frac{1-q_{j-1}}{q_{j-1}}\right)^{|(S-S_{\vec{k}})\cap R_j
|}
\left(\frac{q_j}{1-q_j}\right)^{|(S_{\vec{k}}-S)\cap R_j|}\right)\nonumber \\
 \leq \prod_{j=1}^{t}\left( P_{R_j}(S\cap R_j)
\left(\frac{1-q_{j-1}}{q_{j-1}}
\frac{q_j}{1-q_j}\right)^{n_j}\right)
\\
 \leq \displaystyle\prod_{j=1}^{t} P_{R_j}(S\cap R_j)
\displaystyle\prod_{j=1}^{t} \left(\frac{1-q_{j-1}}{q_{j-1}}
\frac{q_i}{1-q_j}\right)^{\left\lceil\frac{n}{t}\right\rceil}\\
=
P_R(S)\left(\frac{1-p_{min}}{p_{min}}\frac{p_{max}}{1-p_{max}}\right)^{\left\lceil\frac{n}{t}\right\rceil}\nonumber
\end{multline}
\end{small}
(\ref{secondo2})  implies that
\begin{small}
\noindent \begin{multline} \sum_{S\subseteq
\sF}P_R(S)\chi_{\epsilon, l}(S)\\
 \leq  2^{\lambda(n,(1-\delta)p_{min}, \epsilon, l,t)}
\sum_{\vec{k} \in A}\sum_{S\in \sF_{\vec{k}}}P_R(S)
\left(\frac{1-p_{min}}{p_{min}}\frac{p_{max}}{1-p_{max}}\right)^{\left\lceil\frac{n}{t}\right\rceil}\\
 =  2^{\lambda(n,(1-\delta)p_{min}, \epsilon, l,t)}
\left(\frac{1-p_{min}}{p_{min}}\frac{p_{max}}{1-p_{max}}\right)^{\left\lceil\frac{n}{t}\right\rceil}\\
\cdot \sum_{S\in \sF}P_R(S)  \leq  2^{\lambda(n,(1-\delta)p_{min},
\epsilon,
l,t)}\left(\frac{1-p_{min}}{p_{min}}\frac{p_{max}}{1-p_{max}}\right)^{\left\lceil\frac{n}{t}\right\rceil}\label{tre3}
\end{multline}
\end{small}
 Now we provide an upper bound on

\[\lambda(n,(1-\delta)p_{min}, \epsilon, l,t)=(f(l)- g(l))n +
\log(4n) + t\]  where
\[
 f(x)=\frac{1-\epsilon}{\pi x^2}\log
\frac{64 e \pi x^6}{1-\epsilon}
\]
and \[
 g(x)=-(1-\delta)p_{min}\log e
\left(1-(1-\epsilon)\left(1+\frac{1}{2x^2}+\frac{1}{\sqrt{2}x}\right)^2\right)
\]
Function $f(x)$ is decreasing for $x\geq
\frac{5(1-\epsilon)^{\frac{1}{6}}}{(1-\delta) p_{min}\epsilon} $,
so it holds that \[f(l)\leq\frac{(1-\epsilon)^{\frac{2}{3}}}{\pi
}\left(\frac{(1-\delta) p_{min} \epsilon}{5}\right)^2\log\left( 64
e \pi \left(\frac{5}{(1-\delta) p_{min}
\epsilon}\right)^6\right)\] Moreover, for every $a, c>0$, it holds
that  $a\log \frac{c}{a}\leq \frac{c}{e}\log e$. Thus, by setting
$a=\frac{\delta p_{min} \epsilon}{5}$ and $c=(64 e \pi
)^{\frac{1}{6}}$, we get
\begin{equation}\label{funf}
f(l)\leq\frac{12}{5(e\pi)^{\frac{5}{6}}}(1-\delta)p_{min} \epsilon
\log e
\end{equation}
Function $g(x)$ is increasing for $x>1$ and, by  a simple
calculus, we obtain
\begin{equation}\label{fung}
 g(l)\geq \frac{5-\sqrt{2}-1}{5}(1-\delta) p_{min} \epsilon \log e
\end{equation}
From (\ref{funf}) and (\ref{fung}),  we obtain
$$\lambda(n,(1-\delta)p_{min}, \epsilon, l,t)\leq - \frac{7}{100}(1-\delta) p_{min} \epsilon n\log e +\log(4n) +t$$
\noindent Moreover since
\[ \left\lceil\frac{n}{t}\right\rceil\log
\left(\frac{1-p_{min}}{p_{min}}\frac{p_{max}}{1-p_{max}}\right)<\frac{6}{100}(1-\delta)p_{min}n\epsilon
\]  (\ref{tre3})  implies
\begin{eqnarray}\label{quattro} \sum_{S\subseteq
\sF}P_R(S)\chi_{\epsilon, l}(S) &\leq& 2^{-
\frac{1}{100}(1-\delta) p_{min} \epsilon n\log e + \log(4n) + t}
\end{eqnarray}
 We now  give an upper bound on the second addend
 in (\ref{due}). Let $\mu=\min\{\mu_1,\mu_2,\ldots, \mu_t\}\geq
p_{min}\lfloor\frac{n}{t}\rfloor$, then

\begin{small} \begin{eqnarray}
 \sum_{S\subseteq 2^R -\sF}P_R(S)\chi_{\epsilon, l}(S)   & \leq & \sum_{S\subseteq 2^R
 -\sF}P_R(S)\nonumber\\
&  = &  1- \sum_{S\subseteq \sF}P_R(S) \nonumber\\
&  =  & 1- \prod_{j=1}^{t}P_R(|S\cap R_j|\geq (1-\delta)\mu_j)\nonumber\\
& =  & 1- \prod_{j=1}^{t}\left(1- P_R(|S\cap R_j|< (1-\delta)\mu_j)\right) \nonumber\\
& \leq &  1-
\prod_{j=1}^{t}\left(1-e^{-\frac{\delta^2}{2}\mu_j}\right)
\nonumber\\
 & \leq  & 1- \prod_{j=1}^{t}\left(1-
  e^{-\frac{\delta^2}{2}\mu}\right)\nonumber\\
 & = & 1- \left(1- e^{-\frac{\delta^2}{2}\mu}\right)^t\nonumber\\
 & \leq &  t e^{-\frac{\delta^2}{2}\mu}e^{t e^{-\frac{\delta^2}{2}\mu}    } \nonumber \\
 & =  &   e^{  \ln t -\frac{\delta^2}{2}\mu + t e^{-\frac{\delta^2}{2}\mu
 }} \nonumber \\
& \leq &   e^{ t -\frac{\delta^2}{2}\mu}\
\nonumber\\
& \leq &  e^{
-\frac{\delta^2}{2}p_{min}\left\lfloor\frac{n}{t}\right\rfloor+ t}
\label{tre}
\end{eqnarray} \end{small}

\noindent Finally, by  combining   (\ref{due}), (\ref{quattro})
and (\ref{tre}),    the theorem follows. \eproof

\section{An almost optimal solution}\label{apx::uppbound}
We now provide an
efficient construction of a covering range assignment for a random
set $S\subseteq R$ of energy cost very close to the  lower bound
$(1-\epsilon) n/\pi$. Then we will transform it, with additional
cost $o(n)$ only, into a feasible broadcast range assignment that
uses $\Theta(\log(n/\log n))$ ranges and such that the (positive)
smallest among  them, i.e. $l_1$, is $\Theta(\sqrt{\log n})$.

\begin{figure}[ht]
\center
\includegraphics[scale=.8]{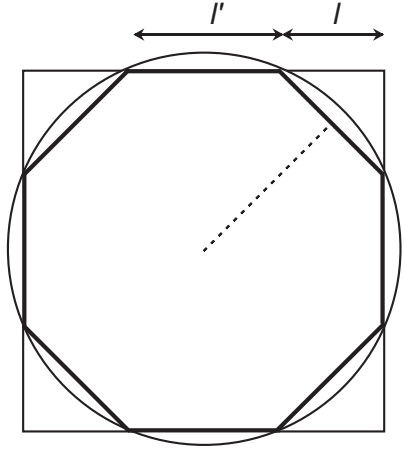}
\caption{The starting configuration of the partition of the square
into triangles and an octagon.} \label{fig.start1}
\end{figure}

\begin{figure}[ht]
\center
\includegraphics[scale=.5]{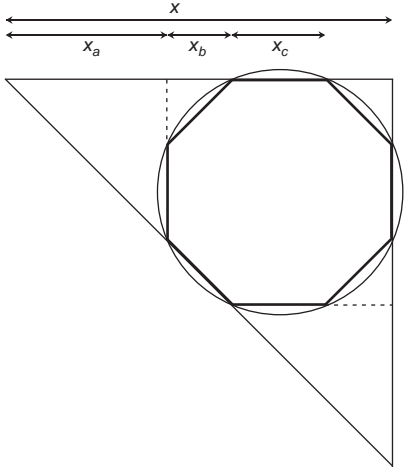}
\caption{The general step of the partition procedure of the square
into triangles and octagons.} \label{fig.step1}
\end{figure}

 \paragraph{The disk covering construction}
 The construction of the covering is recursive and exploits a
tiling of the square with octagons and triangles.
\\  The square $R$ of side $\sqrt{n}$ is partitioned into
four triangles and an octagon (see Figure \ref{fig.start1}); up to
when there exists a triangle with side $c \sqrt{\log n}$, it is
further on partitioned into five triangles (three small and two
big triangles) and an octagon (see Figure \ref{fig.step1}).

\noindent
Starting from this partition, it is possible to produce a
\emph{disk} covering $\OCT$ as follows (in the sequel, a range
assignment  is seen as     a  \emph{disk} assignment with centers
on nodes in $S$):

\begin{itemize}
\item for each triangle of the partition, if it contains at least
one node, then  one of them is selected as    center of a disk
having radius $c\sqrt{2\log n}$. Observe that this disk covers any
other point inside the same triangle.

\item for each octagon, if it contains at least one point that is
not covered yet, then    Lemma  \ref{lemma::threshconn} implies
that
   there is   a node at distance   $O(\sqrt{\log n})$
to the center of the octagon, w.h.p. Let this node be the center
of a disk having radius $r + O(\sqrt{\log n})$, where $r$ is the
radius of the disk that circumscribes the considered octagon; the
introduced disk covers all   points in the octagon.
\end{itemize}
\begin{theorem} Given   a random set $S$,   then, w.h.p.,   disk
covering $\OCT$ has cost
\[\mis(\OCT) \leq 1,12\frac{n}{\pi}+ o(n)\]
\end{theorem}
\proof Let $S$ be the set of all the octagons in the partition,
and for each $s\in S$ call $r_s$ the radius of the disk that
circumscribes octagon $s$. Denoting by $t$ be the number of
triangles in the partition,     by construction it holds that:
\begin{equation}\label{upper1}
\mis(\sC)\leq  2tc^2\log n + \sum_ {s\in S}(r_s + c\sqrt{\log
n})^2
\end{equation}
Let $l$ be the side of the triangles created during the first step
and let $l'$ be the side of the first octagon (see Figure
\ref{fig.start1}). The following equations hold:  $2l+l'=\sqrt{n}$
and $l'=\sqrt{2}l$. From these, we derive:
\begin{equation}\label{inizio}
\begin{array}{ccc}
l=\frac{\sqrt{n}}{2+\sqrt{2}}  &\quad &
l'=\frac{\sqrt{2n}}{2+\sqrt{2}}\end{array}
\end{equation}
The recursive step depicted in Figure \ref{fig.step1}
produces triangles of two different sides and an octagon. Let
$x_a, x_b$ and $x_c$ be the lengths of the sides of the bigger
triangles, the smaller triangles and the octagon, respectively.
These lengths are tied from the following relationships:
$x_c=\sqrt{2}x_b$, $x_a=x_b+x_c$ and  $x_a+2x_b +x_c=l$, implying:
\begin{equation}\label{gadget}
\begin{array}{ccc}
x_a=\frac{l}{\sqrt{2}+1}   &
x_b=\frac{l}{(\sqrt{2}+1)^2}  &
x_c=\frac{\sqrt{2}l}{(\sqrt{2}+1)^2}
\end{array}
\end{equation}
From    (\ref{inizio}) and (\ref{gadget}), the triangles of the
partition,   generated during step $i$, have side length
$x_i=\frac{\sqrt{2n}}{(2+\sqrt{2})(\sqrt{2}+1)^i}$, where $0\leq
i<k$ and $k$ is the smallest integer value such that
$x_i<c\sqrt{\log n}$, i.e.,
\begin{equation}\label{kbound}
k=\left\lceil \frac{1}{2} \log_{(1+\sqrt{2})} \frac{n}{c^2
(2+\sqrt{2})^2\log n} \right\rceil
\end{equation}
Observe that  all   octagons (but the first one) of the partition
are produced by partitioning some triangle of side length $x_i$,
$0\leq i < k$. Denote by  $r$ the radius of the disk that
circumscribes the first octagon,  by $r_i$ the radius of the disk
that circumscribes the octagon produced by partitioning a triangle
of side length $x_i$, and by  $t_i$ the number of such triangles.
Then, we can rewrite   (\ref{upper1}) as follows:
  \begin{multline}
    \mis(\sC) =  t\cdot o(\log n) +
    \sum_{j=0}^{q-1}\sum_{i=0}^{k-1}t_i(r^2_i+ 2c\sqrt{\log n} r_i +
    c^2\log n) \\  + (r+c\log n)^2\label{upper2}
  \end{multline}
  We remind that the radius of the disk that circumscribes
a regular octagon having side  $l$ is
$\frac{l}{\sqrt{2-\sqrt{2}}}$. So, we can use   (\ref{inizio}) and
(\ref{gadget}) to compute the following values of $r$ and $r_i$,
respectively, where $0\leq i <k$:
\begin{equation}\label{raggio}
\begin{array}{cc}
 r=\frac{\sqrt{2n}}{(2+\sqrt{2})\sqrt{2-\sqrt{2}})} &r_i=
  \sqrt{\frac{2n}{2-\sqrt{2}}}\frac{1}{(\sqrt{2}+1)^{i+2}}\frac{1}{2+\sqrt{2}}
 \end{array}
\end{equation}
 In order to compute the value of $t_i$, observe that
trivially $t_0=4$ (see Figure \ref{fig.start1}) and $t_1=8$ (see
Figure \ref{fig.step1}).   At step $i$, $t_i=2t_{i-1}+3t_{i-2}$.
Unrolling the recursion we get:
\begin{equation}\label{numtri}
t_i=3^{i+1} + (-1)^i\leq 3^{i+1} + 1
\end{equation}
In order to evaluate $\mis(\sC)$, we bound all terms appearing in
(\ref{upper2}) by  exploiting (\ref{raggio}) and (\ref{numtri}):
\begin{eqnarray}\label{quadrato}
\sum_{i=0}^{k-1}t_ir^2_i & <& \frac{n}{(2+\sqrt{2})(\sqrt{2}+1)^4}
\sum_{i=0}^{+\infty}\frac{3^{i+1}+1}{(\sqrt{2}+1)^{2i}}
\nonumber\\
&=&\frac{n}{(2+\sqrt{2})(\sqrt{2}+1)^4}\frac{(\sqrt{2}+1)(4\sqrt{2}+3)}{2\sqrt{2}}
\nonumber\\
&<& \frac{(64-45\sqrt{2})n}{4\sqrt{2}}
\end{eqnarray}
\begin{eqnarray}\label{quadratoiniz}
(r+c\log n)^2 = \frac{(2-\sqrt{2})n}{2} + o(\sqrt{n\log n})
\end{eqnarray}
By combining Equations (\ref{numtri}) and (\ref{kbound}) we
obtain:
\begin{eqnarray}
t&=&3^{k+1}+1 \nonumber \\
&\leq&
9\left(\frac{1}{c(2+\sqrt{2})}\right)^{\frac{1}{\log_3(\sqrt{2}+1)}}
\left(\frac{n}{\log
    n}\right)^{\frac{1}{2\log_3(\sqrt{2}+1)}}\nonumber\\
 &=& o\left(\frac{n}{\log n}\right)\label{tbound}
\end{eqnarray}
where the last step is true because
$\frac{1}{2\log_3(\sqrt{2}+1)}<1$. Furthermore,
\begin{equation}\label{lineare}
2c\sqrt{\log n}\sum_{i=0}^{k-1}t_ir_i  = o(n)
\end{equation}
Equation (\ref{numtri}) implies that $\sum_{i=0}^{k-1}t_i<t^k$;
Then, from    (\ref{tbound}) we get:
\begin{equation}\label{costante}
c^2\log n\sum_{i=0}^{k-1}t_i =o(n)
\end{equation}
By combining formulas (\ref{upper2}), (\ref{quadrato}),
(\ref{quadratoiniz}), (\ref{lineare}) and (\ref{costante}) we
conclude that
\[ \mis(\sC)= \frac{(64-45\sqrt{2})n}{4\sqrt{2}} +
\frac{(2-\sqrt{2})n}{2} + o(n)<1.1204\frac{n}{\pi} +o(n)\] \eproof

\begin{figure}[ht]
\center
\includegraphics[scale=1]{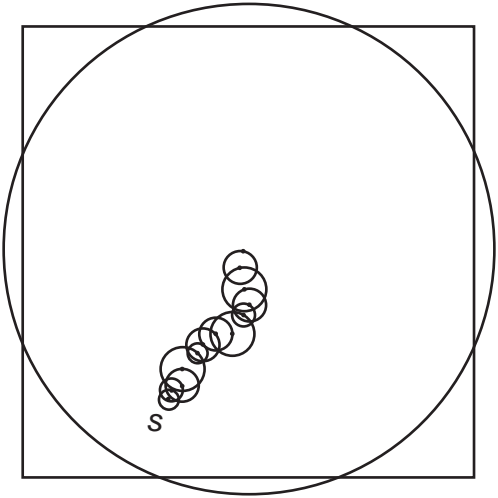}
\caption{Construction of the chain of disks connecting the source
to the center of the first disk.} \label{fig.path}
\end{figure}

\paragraph{From Covering to Broadcasting}
 In order to
guarantee that the produced covering becomes a broadcast, we need
to connect  the source to  the disk centers in   $\OCT$. We start
from the source, located in any place of the square, and build a
chain of disks towards the center of the grid. Thanks to
   Lemma \ref{lemma::threshconn},  the maximum radius of such disks   can be bounded
   by  $O(\sqrt{\log
n})$, w.h.p. (see Fig. \ref{fig.path}). We now show that the
additional cost due to this construction turns out to be
sub-linear.

\noindent The cost of the connection between the source and the
center of the square is   $O(\sqrt{n} \sqrt{\log n})$ w.h.p. Then
we have to connect all the other centers to points already reached
by the information sent from the source. The total cost due to
this step is bounded by $\sum_{j=1}^k t_j x_j O(\sqrt{\log n})$.
By replacing the formulas for $t_j$ and $x_j$ we get:
\begin{multline*}
\sum_{j=1}^k t_j x_j O(\sqrt{\log n}) = O(\sqrt{\log n})
 \sum_{j=1}^k (3^{j+1}+1) (\sqrt{2}-1)^j l_0\\
 =  O(\sqrt{\log n})
 {{\sqrt{n}} \over {2+\sqrt{2}}}(\sum_{j=1}^k 3^{j+1}(\sqrt{2}+1)^j
 + \sum_{j=1}^k (\sqrt{2}-1)^j)\\
= O(\sqrt{\log n}) {{\sqrt{n}} \over
 {2+\sqrt{2}}} \Theta((3(\sqrt{2}-1))^{k+1}) \\
= O(\sqrt{\log n})
 {{\sqrt{n}} \over {2+\sqrt{2}}} ({{1} \over {c(2+ \sqrt{2})} }
 \sqrt{{n} \over {\log n}})^{{1} \over {\log_{3(\sqrt{2}-1)}
 (\sqrt{2}+1)}}
\end{multline*}
This cost is sub-linear since  it is
 $O(n^{0.63})$.
 It is not hard to verify that the above overall construction can
 be performed in $O(n\log n)$ time.

\section{An efficient distributed   protocol}\label{sec::distrib}
Let us consider the following   simple algorithm to construct a
broadcast range assignment. Let $l$ be \emph{any} range in
$\Gamma$ such that $l \ge 2\sqrt{2} c \sqrt{\log n}$ where $c$ is
the constant determined by Lemma \ref{lemma::threshconn} below.

\noindent \textbf{Algorithm \cellbr .}
\begin{small}
\begin{description}
 \item[a.] Grid $R$ is partitioned into  \emph{square
cells} of side length $\lambda = l/(2\sqrt{2})$.

\item[b.] In every non-empty cell, choose one of  its nodes and
assign range $l$ to it. This node is  called the \emph{pivot} of
the cell.

\item[c.] The cell containing the source will have the source as
pivot.

\item[d.] All other nodes have range 0.

\end{description}
\end{small}


  \noindent
  The
proof of the following lemma is a simple application of Chernoff's
Bound.

\begin{lemma}\label{lemma::threshconn}
Let $ p_{min}$,   $ p_{max}$, and $c$  be  three   constants such
that
   $0< p_{min}\leq p_{max}<1$ and $c \ge 16/p_{min}$. Let $S\subseteq R$
be a random grid. Consider the partition of $R$ into square
\emph{cells} of side length $\lambda$  where $ c \sqrt{\log n} \le
\lambda \le
  \sqrt n$.
Then, a constant $\gamma > 0$ exists such that   every cell
contains w.h.p. at least $\gamma \lambda ^2$ nodes. Constant
$\gamma$ can be set as $(1/2)p_{min}$.
\end{lemma}

\noindent
It is then  easy to prove the following

\begin{theorem}\label{thm::cellbro}
Algorithm \cellbr\ yields a  broadcast    range assignment $r$
that is w.h.p. feasible and  its cost satisfies

\[ \mis(r) = \frac {n}{\lambda^2} \cdot (2\sqrt{2}
\lambda)^2 = 8 n \] \end{theorem}

\noindent Thanks to our lower bound in Theorem \ref{t2}, \cellbr\
yields \emph{constant} approximation.

\paragraph{Making it in distributed way}  Algorithm \cellbr\ can be
converted, \emph{without paying any extra energy cost}, into an
efficient, \emph{energy-load balanced} protocol that performs a
\emph{sequence of broadcast operations}.  We describe the protocol
for the many-broadcast scenario and, thus,
 besides minimizing the
energy spent by a single  broadcast operation, we   aim to evenly
distribute the transmission task  among all nodes (but the
source).

\noindent According to the standard radio communication model
\cite{BGI92,CGR02,KKP01}, we assume that nodes act in discrete
uniform \emph{time steps} and are non spontaneous. However, we
assume a weaker, \emph{local} synchronous model:  if, at a given
time step $t$,  the range of a message transmission covers a cell,
then, at time step $t+1$, (only)  the nodes  of that cell are
activated
  and, so,  they will agree on the same time step.
  We assume that every node $v$ knows the number $n$ of points and  its
relative coordinates in the square grid $R$. From its relative
coordinates every node computes a unique \emph{local} label with
respect to its cell. These local labels vary from $1$ to
$\lambda^2$.\\
 The $k$-th
    message sent  by  the source    is  denoted as $\mbox{\sc{m}}_k$.   \emph{Phase}  $k$
     consists of the sequence of time steps where   $\mbox{\sc{m}}_k$ is
     broadcasted. We assume that $\mbox{\sc{m}}_k$
     contains the value $k$.

\noindent
 The protocol performs,   in parallel, two tasks: \emph{i)}
it constructs a broadcast communication graph starting from the
source and \emph{ii)}   transmits the source message along this
graph to all nodes.
  The    procedure is executed for
every broadcast operation from source $s$. Every node keeps a
local counter $\conta$ initially set to $-1$.

\noindent
\textbf{Procedure \bro ($\mbox{\sc{m}}_k$)} 

\begin{small}

\noindent \textbf{Source $s$}     transmits, with range $l$,
$\langle \mbox{\sc{m}}_k, i\rangle$ where $i$ is  the index of its
cell.

 \noindent
 \textbf{All nodes (but $s$):}  
\begin{itemize}

\item If $(k \le \gamma \lambda^2)$ then  \hspace{1cm} ($\gamma$
is the constant of Lemma \ref{lemma::threshconn})
    \begin{itemize}
                \item When a node $v$ receives, for the first
                time w.r.t.  phase $k$,
$\langle \mbox{\sc{m}}_k, i\rangle$  from the pivot of a neighbor
cell $i$, it becomes active.
        \item An active node, at every time step,  increments  its  local counter $\conta$
 by one and checks whether its local label is
equal to the value of its
  $\conta$. If this is the case, it becomes the pivot of its
cell and transmits, with range $l$,   $\langle \mbox{\sc{m}}_k,
j\rangle$ where $j$ is  the index of its cell.

                \item When an active node in cell $i$  receives $\langle \mbox{\sc{m}}_k, i\rangle$,
                  it (so the pivot as well)
records in a \emph{local array}  $P[k]$  the current value of its
  $\conta$, i.e. the local label of the pivot, and becomes
inactive.

         \end{itemize}
    \item else (i.e. $(k >  \gamma \lambda^2)$)
    \begin{itemize}
        \item When a node $v$ receives, for the first time w.r.t. phase $k$,
$\langle \mbox{\sc{m}}_k, i\rangle$  from the pivot of a neighbor
cell $i$, it checks if its local label is equal to $P[k
\bmod{\gamma \lambda^2}]$. If this is the case, it becomes the
pivot of its cell and transmits, with range $l$,  $\langle
\mbox{\sc{m}}_k, j\rangle$ where $j$ is the index of its cell.
    \end{itemize}

\end{itemize}

\end{small}

\begin{fact}\label{fact::properties}  Even though nodes  initially  do not know anything about each other,
 all nodes in the same cell are activated (and disactivated) at the same
time step; so, their local counters share the same value at every
time step. Furthermore, after the first $\gamma \lambda^2$
broadcast operations (i.e. phases), all nodes in the same cell
know the set $P$ of pivots of that cell.
\end{fact}

\noindent More precisely, if $j_0 < j_1< \ j_2< \ldots j_k \ldots$
are the local labels of the nodes in a cell, then, during the
first $\gamma \lambda^2$ broadcast operations (i.e. phases), the
pivot of the cell at phase  $k$ will be the node having local
label $j_k$.

\noindent Procedure  \bro\  has the following properties.

  \noindent
\textbf{Energy Cost.}   As for  each single  broadcast operation,
\bro\ yields a broadcast range assignment equivalent to that of
\cellbr . So, Theorem \ref{thm::cellbro} holds  as well.

 \noindent \textbf{Work Complexity.}
\begin{definition} \label{def:workcompl}
Let $\{\mbox{\sc{g}}_1,\mbox{\sc{g}}_2, \ldots, \mbox{\sc{g}}_h
\}$ be the set of  all   messages sent  by the nodes according to
  a protocol $P$. Then, the \emph{work complexity} of $P$ is
\[  \sum_{i=1}^{h} l_i^2, \  \mbox{ where } \ l_i \ \mbox{ is the range used to send } \ \mbox{\sc{g}}_i  \]
\end{definition}
 The overall number of node transmissions (i.e. the
message complexity)  of every execution of \bro\   is $8 n/l^2$.
Each transmission has range $l$, so the work complexity  is not
larger than $8n$. \\ As for the many-broadcast scenario, our lower
bound in Theorem \ref{t2} easily implies that a work
$k(1-\epsilon)(n/\pi)$ is w.h.p. required to perform a sequence of
$k$ broadcasts (since the lower bound holds for the energy cost).
It follows that our protocol achieves an almost optimal work
complexity for the many-broadcast operation as well.

  \noindent \textbf{Load Balancing and Network Lifetime.} The \emph{expensive} pivot's
task is evenly assigned, w.h.p., to $\gamma \lambda^2$
   nodes (see Lemma
\ref{lemma::threshconn}) in the same cell by using a  round robin
schedule.
 This
 is crucial when the number of broadcasts increases and nodes have limited battery charge.
As for the many-broadcast operation, it is possible to show that
our protocol achieves an almost maximal lifetime according to the
  consumption  model in \cite{CKOZ06,CCFS07}.  In this model,
 the goal is to maximize the  lifetime of the
network while
 guaranteeing, at any  phase $k$,  a broadcast operation from the  source.
 Formally, each node  $v$ is initially equipped with a battery charge\footnote{Here we assume that,
 at the very beginning, all nodes
 are in the same energy situation.}  $B>0$.   Whenever a node
 transmits with range $l$, its battery charge is
  reduced by amount $\beta \cdot l^2$ where
 $l$ denotes the range assigned to node $v$  and $\beta>0$ is a fixed constant
 depending on the adopted technology.
We assume $\beta =1$, however, all our results  holds for any
$\beta>0$. \\ Then, the  \maxlife\  problem  is to maximize the
number of independent broadcast operations till some node will die
(i.e. its battery charge becomes 0).
 In \cite{CKOZ06},
\maxlife\ is shown to be  $\np$-hard.

\begin{theorem}
   \bro\  performs a sequence of independent broadcast
operations whose length is  only a constant factor smaller than
the optimum, w.h.p.
\end{theorem}

\sketchofproof We have already observed that the work complexity
of  \bro\ for  any single broadcast operation is not larger than
$\alpha \opt$, where $\alpha$ is a positive constant and $\opt$ is
the optimal work complexity.  So, the maximal number of
independent broadcast operations is not larger than $nB/\opt$.
Thanks to the local round robin strategy in every cell, the energy
load of the many-broadcast operation is well balanced  over at
least  a (large) constant fraction $\eta$ of all nodes.  So the
number of broadcast operations   perform by \bro\ is at least
$\frac{\eta n B }{\alpha \opt} = (\eta/\alpha) \frac{nB}{\opt}$,
w.h.p. \eproof

\noindent \textbf{(Amortized) Completion Time.}
\begin{theorem} \label{thm::timebound}
  The amortized completion time (i.e. the average number of time steps to perform one broadcast
  operation)   over a sequence of $T$ broadcast operations
    is  w.h.p.
\[O(l\sqrt n /T
  +   \sqrt n/l) \]
\end{theorem}
\sketchofproof  For a single broadcast operation performed by
\bro, we define the \emph{delay} of a cell  as the number of time
steps from  its activation time till the selection of its pivot.
 Observe that the sum of delays introduced by a cell during
 the first $\gamma \lambda^2 $ broadcasts is at most $\lambda^2=\Theta(l^2)$.
Then,  the delay of any cell becomes 0 for all broadcasts
 after the first  $\gamma \lambda^2$ ones. Moreover,
  a broadcast can pass over at most $O(\sqrt n /l)$
cells.   By assuming that a maximal length path
 (this length being  $\Theta(\sqrt n /l))$
  together with maximal cell delay  can be found in each of the first
 $\min \{ \gamma\lambda^2 , T \}$ broadcasts,    we can bound the maximal overall delay with
\begin{equation} \label{Eq::b1}
O( l \sqrt n )
\end{equation}

\noindent Finally, the number of time steps required by every
broadcast without    delays
 is
 \begin{equation} \label{Eq::b3}
 O(\sqrt n/l)  \end{equation} since   the length of any path on the broadcast
tree  is $O(\sqrt n/l)$.
 By combining
(\ref{Eq::b1}) and (\ref{Eq::b3}), we get the theorem bound.
\eproof

\noindent For brevity's sake, the amortized completion time has
been analyzed without considering the \emph{interferences} due to
  \emph{collisions} among   pivot transmissions \cite{BGI92}. However,
  in order to avoid such collisions, we can  further   organize \bro\ into
iterative \emph{stages}: in every stage, only cells with not
colliding pivot transmissions are active. Since the number of
cells that can interfere with a given cell is constant, this
further scheduling will increase the overall time  by a constant
factor only. This iterative process can be efficiently performed
in a distributed way since every node knows $n$ and its position,
so it knows its cell.

\begin{corollary}
The completion time of one single broadcast operation is $O(l
\sqrt n)$.
\end{corollary}

\noindent The worst scenario for our protocol occurs  when $T$ is
small, say  $T = O(1)$. Indeed, assume that a transmission range
$l = \Theta(\sqrt{\log n})$ is available in $\Gamma$, then we get
an amortized completion time $O( \sqrt {n \log n})$ that is a
factor $\log n$ larger then the optimum. Notice that    in this
case, the network
diameter is  $\Theta(\sqrt {n / \log n})$ w.h.p.\\
    Whenever  $T=\Omega(\log
n)$, we instead get $O(\sqrt{n / \log n})$  amortized completion
time which is optimal.

\begin{table*}[t]
\begin{center}
\begin{small}
\begin{tabular}{|c||c|c|c|c||c|c|c|c|}
\cline{2-9}
\multicolumn{1}{c}{}&\multicolumn{8}{|c|}{$\mis(\cellbr)$  vs  $\mis(MST)$}\\
\cline{2-9}
\multicolumn{1}{c}{}&\multicolumn{4}{|c||}{$p=0.2$}&\multicolumn{4}{|c|}{$p=0.5$}\\
\hline
$\sqrt{n}$&\# of feasible sol.& min & average & max&\# of feasible sol.& min & average & max\\
\hline
13 &744/1000&2.044 & 2.072&2.118&1000/1000&2.544&2.808&3.086\\
\hline
20 &892/1000&1.896 &2.092 &2.145&999/1000&2.506&2.666&2.994\\
\hline
25 &762/1000&2.040&2.127 &2.164&998/1000&2.341&2.617&2.808\\
\hline
30 &740/1000&2.163&2.217&2.242&997/1000&2.512&2.610&2.666\\
\hline
50 &858/1000&2.313&2.398&2.450&999/1000&2.673&2.824&2.949\\
\hline
100&967/1000&2.300&2.347&2.352&1000/1000&2.604&2.659&2.702\\
\hline
\end{tabular}
\end{small}
\caption{Experimental results.}
\label{table::connectivity_frequency}
\end{center}
\end{table*}

\subsection{Experimental results} \label{sec::experiments}

In this subsection, we present  the experimental results we have
obtained by running   Algorithm \cellbr. We have generated $1000$
instances for every  side length  \[ \sqrt n \in \{ 13, 20, 25,
30, 50, 100 \} \] and for    \emph{node-probability} $p \in \{
0.2, 0.5 \}$. As usual, our implementation benefits of  some
parameter tuning and optimization: the pivot node (but the source
node) inside every cell is the one closer to the center of the
cell and useless, redundant ranges are removed. These tasks can be
performed also by the distributed protocol, after the first phase
(i.e. for $t \geq \gamma \lambda^2$), without paying any extra
energy cost since, after that time, every node of a cell knows all
its cell neighbors. Moreover, the transmission range   $l$ is
set\footnote{Notice that, for the tested sizes $n$, this range is
smaller than the threshold $2\sqrt 2 c \sqrt{\log n}$ defined in
Section \ref{sec::distrib}: this is the reason why the feasibility
rate is not 100\% for large $n$.} to $\sqrt{2}\log n$, while the
cell-size parameter $\lambda$ is set to $\log n$. Notice that,
according to such choices, the feasibility (i.e., the existence of
a path from the source node to all   other nodes in the induced
communication graph) is tested too. In
Table~\ref{table::connectivity_frequency} (columns "\# of feasible
sol."), the number of feasible solutions for the different
combinations of $n$ and $p$ are reported.

\noindent The solution costs of \cellbr\ are compared to the cost
of the solution returned by the centralized MST-based algorithm.
We remind that while the energy cost of \cellbr\ is an upper bound
on  the work complexity of  our  distributed procedure \bro\, the
energy cost of the MST-based solution does not provide any
information about the work complexity of its distributed
implementations (this  can be much larger).  \\
Table~\ref{table::connectivity_frequency} shows, for all chosen
values of $p$ and $\sqrt n$, the minimum, average and maximum
ratio between the   costs of the solutions returned by the two
algorithms. As for \cellbr ,  only the costs of \emph{feasible}
solutions are considered.

\end{document}